\documentclass[sigconf]{acmart}

\usepackage{graphicx}
\usepackage{booktabs} 
\usepackage{subfigure}
\usepackage{url}
\usepackage{color}
\usepackage{listings}
\usepackage{flushend}
\usepackage{bbding}

\copyrightyear{2021}
\acmYear{2021}
\setcopyright{none}
\acmConference{ICPE'21}{April 19 - 23, 2021}{Rennes, France}

\begin{document}

\title{The Granularity Gap Problem: A Hurdle for Applying Approximate Memory to Complex Data Layout*}

\author{Soramichi Akiyama}
\affiliation{%
  \institution{The University of Tokyo, Tokyo, Japan}
}
\email{akiyama@ci.i.u-tokyo.ac.jp}

\author{Ryota Shioya}
\affiliation{%
  \institution{The University of Tokyo, Tokyo, Japan}
}
\email{shioya@ci.i.u-tokyo.ac.jp}

\begin{abstract}
  The main memory access latency has not much improved for more than two decades while the CPU performance had been exponentially increasing until recently.
  {\it Approximate memory} is a technique to reduce the DRAM access latency in return of losing data integrity.
  It is beneficial for applications that are robust to noisy input and intermediate data such as artificial intelligence, multimedia processing, and graph processing.
  To obtain reasonable outputs from applications on approximate memory, it is crucial to protect critical data while accelerating accesses to non-critical data.
  We refer the minimum size of a continuous memory region that the same error rate is applied in approximate memory to as the {\it approximation granularity}.
  A fundamental limitation of approximate memory is that the approximation granularity is as large as a few kilo bytes.
  However, applications may have critical and non-critical data interleaved with smaller granularity.
  For example, a data structure for graph nodes can have pointers (critical) to neighboring nodes and its score (non-critical, depending on the use-case).
  This data structure cannot be directly mapped to approximate memory due to the gap between the approximation granularity and the granularity of data criticality.
  We refer to this issue as the {\it granularity gap problem}.
  In this paper, we first show that many applications potentially suffer from this problem.
  Then we propose a framework to quantitatively evaluate the performance overhead of a possible method to avoid this problem using known techniques.
  The evaluation results show that the performance overhead is non-negligible compared to expected benefit from approximate memory,
  suggesting that the granularity gap problem is a significant concern.
  \let\thefootnote\relax\footnote{{\bf *This is an extended version of our conference paper published in the 12$^{\rm th}$ ACM/SPEC International Conference on Performance Engineering (ICPE).}}\addtocounter{footnote}{-1}
\end{abstract}
\maketitle

\section{Introduction}
\begin{figure}[!t]
  \centering
  \includegraphics[width=0.9\columnwidth]{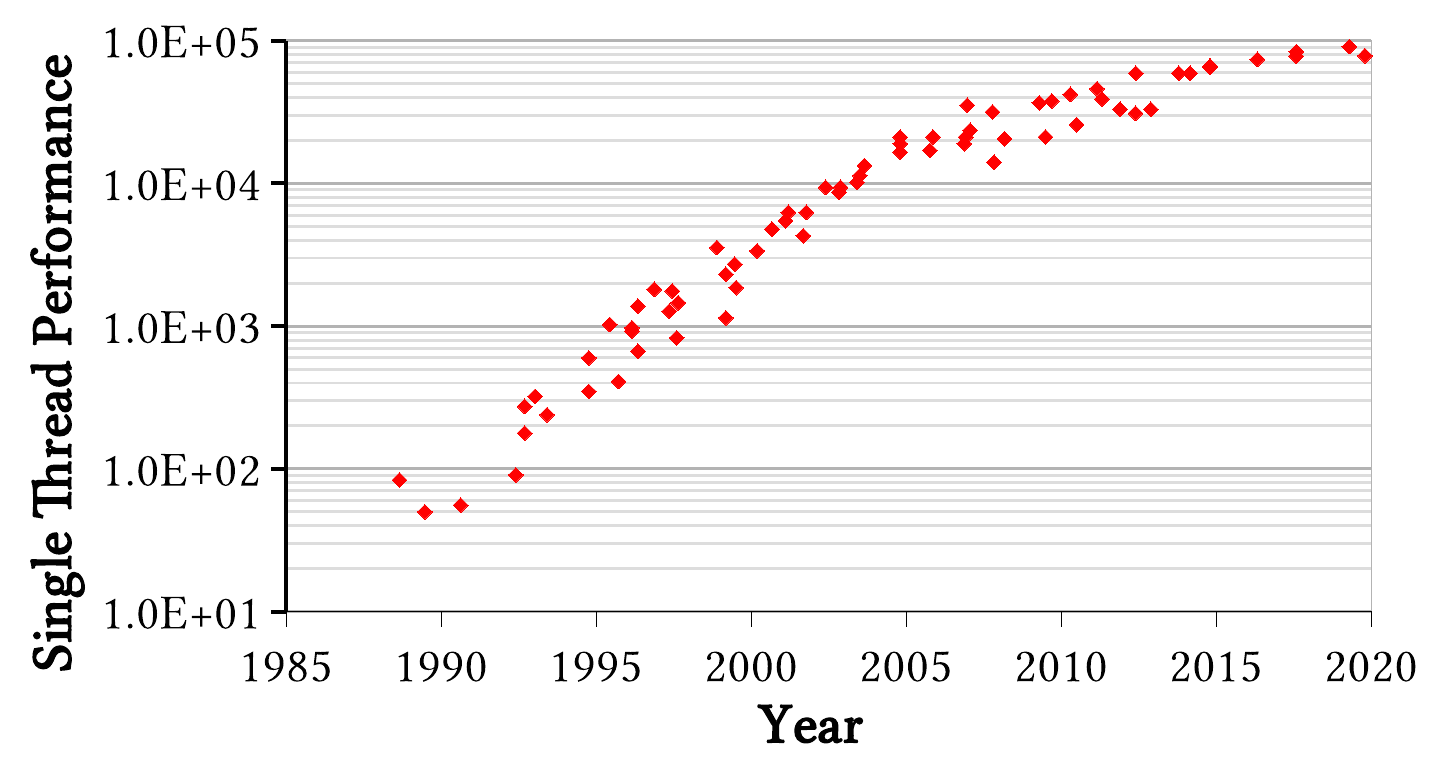}
  \caption{\label{figure:single_thread_performance}Trend of single thread performance over time (normalized to SPEC CPU 2006 score $\times$ 1000): the performance had been increasing exponentially until recent years.}
\end{figure}

The impact of main memory access latency to the overall performance is much larger on a computer today than in the past.
This is because the performance gap between the main memory and the CPU has ever been enlarging.
Figure~\ref{figure:single_thread_performance} shows the single thread performance of server-class CPUs plotted over time\footnote{Data provided in~\cite{karlrupp_github} under CC BY 4.0 (\url{https://creativecommons.org/licenses/by/4.0/}).}
The figure shows an exponential growth of the single thread performance until recent years.
In contrast, the access latency of DRAM that the main memory consists of has been almost the same for more than two decades.
As shown in~\cite{Chang2016}, the speedup of the major latency sources of DRAM over time is very marginal, especially when compared to the exponential growth of the CPU performance.
Because DRAM access latency occupies substantial amount in a random memory access latency\footnote{For example, a random memory access latency in the machine shown in table~\ref{table:env} is around 82 ns (measured by Intel MLC), while the sum of the three major latency sources shown in~\cite{Chang2016} of the DRAM module in this machine is around 45 ns~\cite{JEDEC_DDR4}.},
there is a strong need to reduce the DRAM access latency to catch up with the CPU performance.

{\it Approximate memory} is a technique to reduce the main memory latency by sacrificing its data integrity~\cite{Koppula2020,Tovletoglou2020,Raha2017,Nguyen2020}.
Prior works have proven that DRAM modules used for main memory can be operated much more aggressively than defined in the specifications.
In concrete, the access latency can be reduced by violating the timing constraints of DRAM internal operations at the cost of increased bit-error rate~\cite{Lee2017,Chang2016,Wang2018,Hassan2016,Zhang2016,Choi2015}.
Prior works try not to expose the increased bit-error rate to applications by operating DRAM to the extent that the error rate is still small enough to have zero bit-flips during applications' runtime.
Approximate memory exploits the same idea more aggressively to reduce the memory access latency by leveraging the error robustness of applications themselves.
It is expected to be beneficial for application domains such as deep learning, multimedia processing, graph processing, and big-data analytics
because these applications are known to be robust to bit-flips to some extent~\cite{Nie2018,Chandramoorthy2019,Mahmoud2019,Mahmoud2020,Chen2019}.

To obtain reasonable outputs from applications on approximate memory, it is crucial to protect {\it critical data} while accelerating accesses to {\it non-critical data}.
For example, suppose we want to accelerate a deep learning application on top of approximate memory.
The matrices that express the weights of each layer are non-critical data because it is known that the accuracy of the trained model does not degrade much even when some bit-flips are injected into them~\cite{Chandramoorthy2019,Givaki2020,Koppula2020,Mahmoud2020}.
On the other hand, pointers from one layer of a network to another or the loop counter that counts the number of epochs are non-critical data because they must be protected from bit-flips.
Therefore, we must control the error rate of memory regions depending on the criticality of the stored in them.

A limitation of approximate memory is that the error rate can be controlled only with the granularity of a few kilo bytes due to the internal structure of DRAM chips.
We refer the minimum size of continuous data to which the same error rate must be applied to as {\it approximation granularity}.
The approximation granularity for a given DRAM module is decided by the {\it row size} of the module.
A row is a sequence of data bits inside a DRAM module that are driven simultaneously with the same timing.
Because approximate memory we focus on is based on tweaking the timing of DRAM internal operations, the approximation granularity is equal to the row size.
The row size of a DRAM module is in the range of 512 bytes to a few kilo bytes.
For example, the row size of a module from Micron~\cite{micron_32gb_spec} is 2~KB, meaning that the approximation granularity of this module is also 2~KB.
This stems from the fundamental limitation of modern DRAM that many bits must be driven in parallel to catch up with requests coming from a fast CPU.

The large approximation granularity makes it difficult to gain benefit from approximate memory for applications that have critical and non-critical data interleaved with a smaller granularity (e.g., 8 bytes).
We refer to this problem as the {\it granularity gap problem}.
This can happen when an application manages its data as an array of data structure that has critical members (e.g., pointers) and non-critical members (e.g., numbers whose small divergence do not affect the application's result).
For a concrete example, suppose an application that traverses an array of graph nodes, and each graph node has pointers to its neighboring nodes and a score of it that is robust to bit-flips.
The non-critical data of this application cannot be stored in approximate memory due to
the difference between the approximation granularity and the granularity of interleaving of critical and non-critical data.

In this paper, we show the granularity gap problem is a significant concern in using approximate memory.
In concrete, the contributions of this paper are summarized as follows:
\begin{enumerate}
\item A source code analysis of widely used benchmarks to prove that many applications potentially suffer from the granularity gap problem, extended from our previous work~\cite{Akiyama2020}.
\item A discussion on pros and cons of a memory layout conversion technique in the context of the granularity gap problem.
\item A framework to quantitatively evaluate the negative performance impact of the memory layout conversion technique.
\item Evaluation results of the negative performance impact on widely used benchmarks, which proves the significance of the granularity gap problem in using approximate memory.
\end{enumerate}

The rest of the paper is structured as follows.
Section 2 introduces the background knowledge of how DRAM and approximate memory works.
Section 3 defines the granularity gap problem and the goal of this paper.
Section 4 analyzes the source code of SPEC CPU 2006 and 2017 benchmarks.
Section 5 explains a memory layout conversion technique to avoid the granularity gap problem, and points out why it is not sufficient.
Section 6 describes our simulation framework and gives quantitative evidence that the granularity gap problem is a significant concernt.
Section 7 reviews related work and Section 8 concludes the paper.

\section{Approximate Memory Architecture and Its Limitation}
\subsection{Overview of Approximate Memory}
Approximate memory is a new technology to mitigate the performance gap between main memory and CPUs.
The main idea is to reduce the latency of main memory accesses at a cost of the data integrity by exploiting {\it design margins} that exist in many DRAM chips today.
The CPU may read a slightly different data from what has been written before to the main memory.
A design margin refers to the difference between a design parameter defined in the specification of a device and the actual value which the device can be operated with.
In particular, we focus on the design margin in the timing of internal operations of DRAM.
Even when some wait-time parameters are shortened than the specification, many DRAM chips can read stored data ``almost'' correctly with a few bit-flips (errors) injected to the data~\cite{Chang2016,Kim2018}.
By controlling the timing of internal operations of DRAM, we can trade reduced main memory access latency with increased bit-error rate.

Approximate memory attracts much research interest due to the ever-increasing performance gap between main memory and CPUs.
Chang~{\it et al.}~\cite{Chang2016} measure the relationship between error rates and latency reduction for a large number of commercial DRAM chips.
Tovletoglou~{\it et al.}~\cite{Tovletoglou2020} propose a holistic approach to guarantee the service level agreement of virtual machines running on approximate memory.
Koppula~{\it et al.}~\cite{Koppula2020} re-train deep learning models on approximate memory so that the models can adapt to errors.
Our previous work~\cite{Akiyama2019} estimates effect of approximate memory to realistic applications without simulation by counting the number of DRAM internal operations that incur errors.

Approximate memory is especially beneficial for machine learning, multimedia, and graph processing applications,
all of which incur many memory accesses and are tolerant to noisy data.
For example, Stazi~{\it et al.}~\cite{Stazi2019} show that allocating data in approximate memory for the x264 video encoder can yield acceptable results,
and our previous work~\cite{Akiyama2019} show that a graph-based search algorithm (mcf in SPEC 2006) can yield the same result as error-free execution even when some bit-flips are injected.
Regarding the performance improvement, Koppula~{\it et al.}~\cite{Koppula2020} show 8\% speedup in average for training various DNN models on approximate memory,
and Lee~{\it et al.}~\cite{Lee2017b} show that using Adaptive-Latency DRAM~\cite{Lee2015} for approximate memory
gives 7\% to 12\% speedup in averag for {\it ``32 benchmarks from Stream, SPEC CPU 2006, TPC and GUPS''} (they do not show numbers for each benchmark though).
The performance improvement of a few to 10+ percent is important to these applications because they are typically executed in large scale data centers,
where only a few \% of relative efficiency improvement results in a huge absolute reduction of energy and/or runtime in total.

\subsection{Design Margin: Timing Constraints}
The design margin exploited to realize approximate memory that we focus on is the {\it timing constraints} of DRAM internal operations.
Although there are other types of approximate memory such as approximate flash memory that leverages multiple levels of programming voltages~\cite{Guo2016} and approximate SRAM based on supply voltage scaling~\cite{Chandramoorthy2019,Yang2017,Esmaeilzadeh2012}, we focus on approximate DRAM in this paper.
A DRAM module is operated by the memory controller that issues electric signals referred to as {\it DRAM commands}.
A DRAM command triggers an internal operation of DRAM such as resetting the voltages of wires to the reference voltage.
A {\it timing constraint} refers to the interval between two DRAM commands and we categorize them into two types:
\begin{itemize}
\item {\bf Type 1} specifies the interval that must pass before the next DRAM command is issued.
  They are defined so that the internal operation of DRAM triggered by the previous command is guaranteed to finish before the next command.
\item {\bf Type 2} specifies the interval where the same command must be issued again within that interval.
  This is defined so that the electrical charges inside DRAM does not leak too much by periodically refreshing them.
\end{itemize}
The actual values of the timing constraints for each type of DRAM module (e.g, DDR4-2400) are specified by JEDEC,
which is an organization that publishes DRAM-related specifications.

Relaxing a timing constraint means either shortening or prolonging an interval defined by the specifications (i.e., ``violating'' the specifications).
It reduces the average access latency of DRAM because commands are served faster (by shortening the Type 1 constraints) and it increases the number of useful commands executed (by prolonging the Type 2 constraints).
However, it increases the possibility that bit-flips are injected into the data because there is no guarantee that a DRAM module works flawlessly when the timing constraints are violated.

\begin{figure}[t]
  \centering
  \includegraphics[width=0.8\columnwidth]{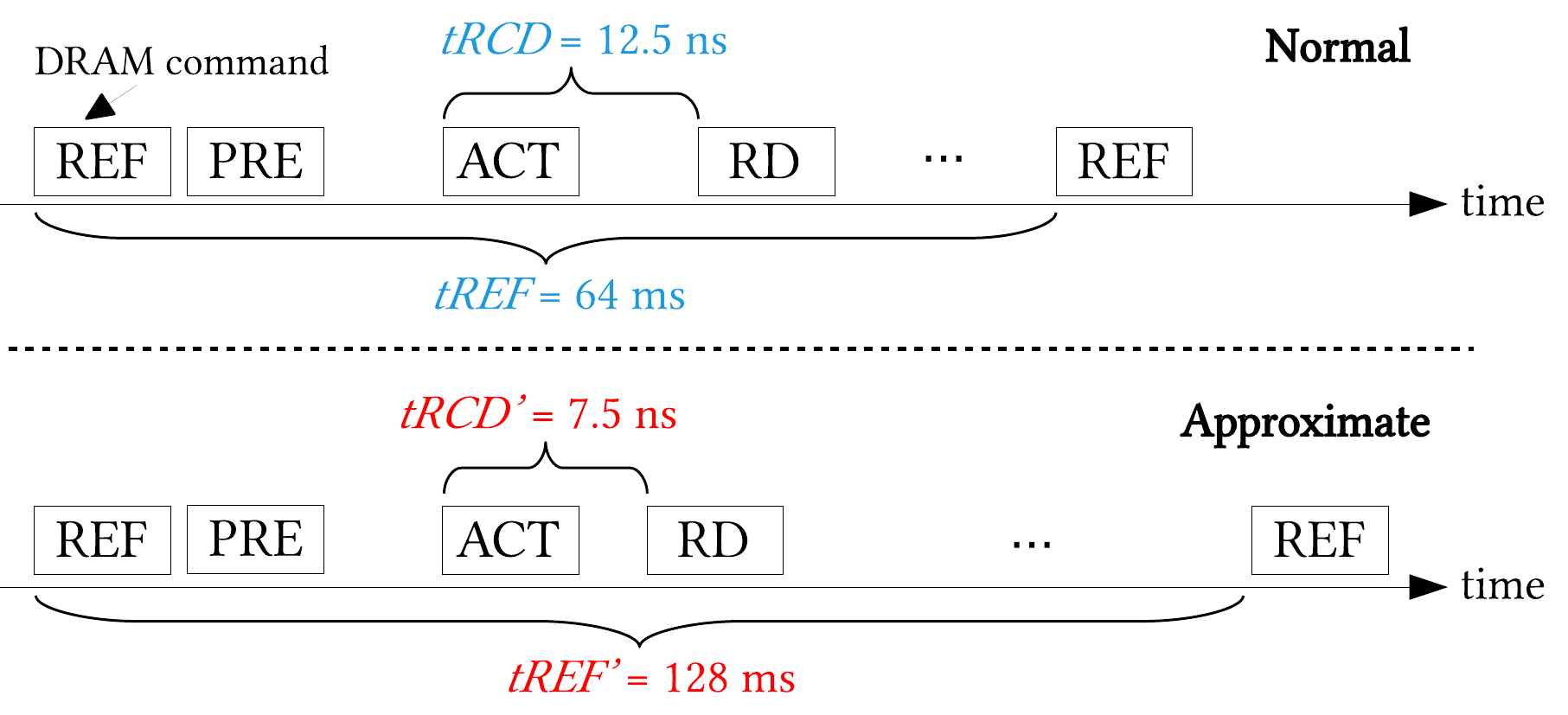}
  \caption{\label{figure:command_timeline}DRAM command sequence of normal memory (top) and approximate memory (bottom): In this example, tRCD is shortened to 7.5~ns and tREF is prolonged to 128~ms, both of which reduce the average latency.}
\end{figure}

Figure~\ref{figure:command_timeline} shows an example of DRAM command sequence in normal memory and approximate memory.
It shows four representative DRAM commands: refresh (\verb|REF|), precharge (\verb|PRE|), activation (\verb|ACT|), and read (\verb|RD|).
In this example, a timing constraint called \verb|tRCD| (Type 1) is shortened from 12.5~ns to 7.5~ns, and one called \verb|tREF| (Type 2) is prolonged from 64~ms to 128~ms.
\verb|tRCD| is the interval that must pass between an \verb|ACT| command and the following \verb|RD| command, and it is around 11 -- 13 ns depending on the DRAM module (e.g., 12.5~ns for DDR3-1600J~\cite{JEDEC_DDR3}).
Chang~{\it et al.}~\cite{Chang2016} found that only a small portion of the data bits experience errors even when \verb|tRCD| is shortened below it.
We explain how an \verb|ACT| command and \verb|tRCD| work inside DRAM more in detail in Section~\ref{section:act_example}.
\verb|tREF| is another timing constraint that specifies the longest interval between two \verb|REF| commands, which refresh DRAM cells to prevent them from losing stored data.
Das~{\it et al.}~\cite{Das2018} and Zhang~{\it et al.}~\cite{Zhang2016} propose to prolong this interval because many DRAM cells can retain data for more than 64~ms in practice.
Because prolonging \verb|tREF| increases the amount of time during which more useful commands are served, it reduces the average DRAM access latency.

\subsection{\label{section:act_example}Closer Look: ACT Command Example}
\begin{figure}[t]
  \centering
  \includegraphics[width=\columnwidth]{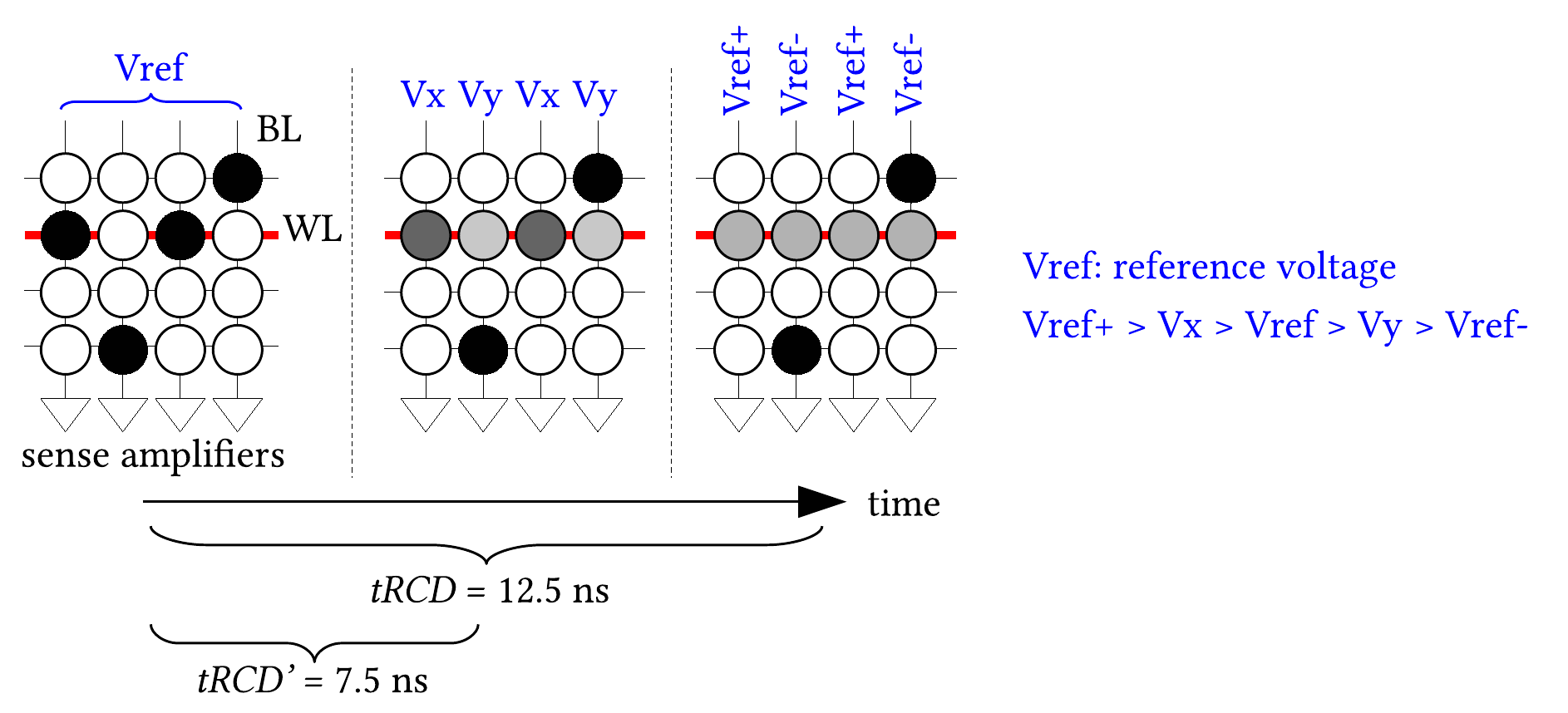}
  \caption{\label{figure:activation_timeline}
    ACT command copies the value of the selected row to the sense amplifieres.
    Left: The BLs are reset to Vref in the reset state.
    Right: After 12.5 ns, the BLs are guaranteed to be either Vref+ or Vref-.
    Middle: If tRCD is reduced to 7.5 ns, the sense amplifiers sense unstable voltages of BLs (Vx and Vy), resulting in a few bit-flips but a shorter latency.
  }
\end{figure}

The left-most side of Figure~\ref{figure:activation_timeline} shows the reset state of DRAM.
The circles show an array of memory cells, where each row is connected by a wordline (WL) and each column is connected by a bitline (BL).
Although a DRAM chip consists of a hierarchy of many of these arrays operated in parallel, we focus on one array here without loss of generality.
A black cell has electric charge in it and a white cell is empty.
A cell with charge in it represents a value of 1, an empty cell represents a value of 0.
In the reset state, the voltages of all the BLs are set to the reference value denoted as \verb|Vref| in the figure.

An \verb|ACT| command takes the target row number as its parameter (for example, the 2$^{\rm nd}$ row from the top).
The WL of the target row is enabled to connect the cells in the target row to the BLs.
The voltages of the BLs connected to cells with charge start being pulled up,
and the voltages of the BLs connected to empty cells start being pulled down.
At the same time, the cells connected to the BLs become intermediate state (denoted by gray circles in the figure) because the capacitance of a BL is much larger than that of a cell.
After \verb|tRCD| (12.5 ns in the figure) has passed, the voltages of the BLs are guaranteed to be either \verb|Vref+| or \verb|Vref-| as shown in the right-most side of the figure.
Finally, the sense amplifiers sense the voltages of the BLs to fetch the values and buffer them.

Although the value of \verb|tRCD| is strictly defined by JEDEC, real DRAM chips are known to have much design margin in this timing parameter.
Previous work~\cite{Chang2016,Kim2019} show that many bits can be fetched correctly even when \verb|tRCD| is shortened by a substantial amount.
The middle of Figure~\ref{figure:activation_timeline} shows how the \verb|ACT| command works when \verb|tRCD| is shortened to 7.5 ns to reduce the memory access latency.
Because 12.5 ns has not yet passed, the voltages of the BLs may have not reached to \verb|Vref+| or \verb|Vref-|,
but they are unstable values denoted as \verb|Vx| and \verb|Vy| in the figure (\verb|Vref+| > \verb|Vx| > \verb|Vref| > \verb|Vy| > \verb|Vref-|).
When the sense amplifiers sense the voltages of the BLs at this point, they may fetch wrong values because the difference of \verb|Vx| and \verb|Vy| against \verb|Vref| are not large enough to sense the values.
Larger the difference of \verb|Vx| and \verb|Vy| against \verb|Vref| is, larger the probability of fetching correct values.
This way, controlling the timing parameter serves as a knob for trading the access latency and bit-error rate.

\subsection{Limitation: Approximation Granularity}
A limitation of approximate memory exploiting design margins in timing constraints is that the {\it approximation granularity} cannot be smaller than a few kilo bytes.
The approximation granularity refers the minimum size of a continuous memory region to which the same error rate must be applied.
This is because the same timing parameter is applied to an entire row as we describe in Section~\ref{section:act_example},
and the size of a DRAM row (also known as a {\it page}~\cite{Jacob2007}) is as large as a few kilo bytes in modern DRAM modules.
This stems from a fundamental constraint that many DRAM cells must be driven in parallel so that slow DRAM can catch up with the high rate of requests coming from the CPU.
Therefore, the same limitation is applicable to DRAM commands other than \verb|ACT| and their timing constraints as well.
For example, refreshing DRAM cells are also done row by row (i.e, an entire row is refreshed at once) and thus prolonging \verb|tREF| also affects an entire row at once~\cite{Jacob2007}.

We give two examples of the row size in real DRAM modules.
A 32 Gb DRAM module from Micron~\cite{micron_32gb_spec} has 64~K rows, 16 banks, and 2 ranks\footnote{Right-most column of Table 3 in~\cite{micron_32gb_spec}.}.
The row size of this module is calculated as:
\begin{equation}
  \frac{32 \rm{Gb}}{64 \rm{K} \times 16 \times 2} = 16 \rm{Kb} = 2 \rm{KB} \label{eq:row_size_micron}
\end{equation}
Note that a column contains more than 1 bit (as explained in p. 416 of~\cite{Jacob2007}), thus multiplying the denominator of the left-most side of equation~(\ref{eq:row_size_micron}) by the number of columns does not match the capacity of the module.
Another 16 Gb DRAM module from SAMSUNG~\cite{samsung_16gb_spec} also has a page size (row size) of 2~KB~\cite{samsung_16gb_spec}\footnote{Right-most column of ``16 Gb Addressing Table'' on page 9 in~\cite{samsung_16gb_spec}.}.
This can be confirmed by a similar calculation to equation~(\ref{eq:row_size_micron}) using the number of rows ($2^{17}$ = 128K) and the number of banks ($2^2$ banks per bank group $\times$ 2 bank groups = 8 banks):
\begin{equation}
  \frac{16 {\rm Gb}}{128 {\rm K} \times 8} = 16 {\rm Kb} = 2 \rm {KB}
\end{equation}

Although this paper focuses only on DRAM, the same limitation is applicable to other memory technologies because the fundamental performance gaps exist between CPUs and any memory technologies known today.
This is also true for non-volatile memory technologies that are emerging to mitigate the energy and density issues of DRAM.
For example, phase change memory (PCM) injects electric pulses to an entire row at once for writing~\cite{Nishi2019}.
If we consider realizing approximate memory with PCM, for example by reducing the length of pulses, the approximation granularity is still limited by its row size.

\section{Critical Data Protection and Challenge}
\subsection{Critical Data Protection}
Even for applications that can tolerate noisy input and intermediate data, they have critical parts of data that must be protected from bit-flips.
For example, deep learning is known to be robust to bit-flips~\cite{Chandramoorthy2019,Givaki2020,Koppula2020,Mahmoud2020} but not all parts of the data are robust to them.
Pointers from one layer of a network to another or the loop counter that counts the number of epochs must be protected from bit-flips.

Protecting critical parts of data requires two steps:
\begin{enumerate}
\item Detecting which parts of data are critical and which parts are non-critical
\item Storing non-critical parts of data into approximate memory while storing the critical parts to normal memory
\end{enumerate}
For step (1), there have been much effort~\cite{Akiyama2019,Ashraf2015,Wei2014,Nie2018,Mahmoud2019} and it is out of the scope of this work,
so we assume that discrimination of critical and non-critical data is given.
For step (2), because the timing constraints are controlled per row,
we must map the critical and non-critical parts of data into different DRAM rows running with different timing parameters:
the timing parameter same as defined in the specification and the one shortened for faster accesses.

\begin{figure}[t]
  \centering
  \includegraphics[width=\columnwidth,bb=0 0 928 296]{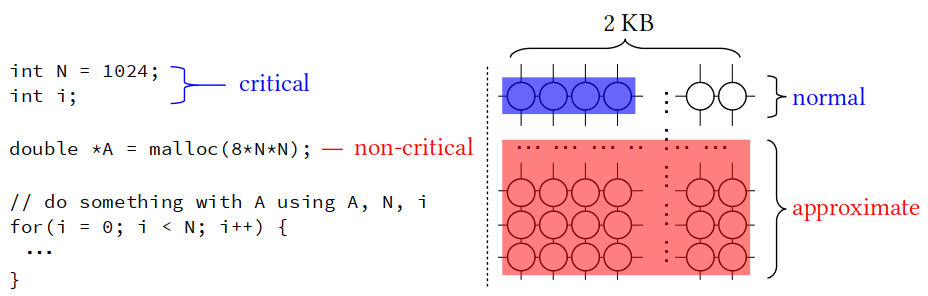}
  \caption{\label{figure:data_partitioning}
    Mapping critical and non-critical parts of data into different DRAM rows
    to protect the former while reducing access latency to the latter.}
\end{figure}

Figure~\ref{figure:data_partitioning} depicts an example of mapping critical and non-critical data into different DRAM rows.
In the figure, suppose the variables \verb|N| and \verb|i| are critical because the former decides the size of allocated memory and the latter is a loop counter,
and the memory region pointed to by \verb|A| is non-critical.
\verb|N| and \verb|i| are mapped to the first row in the figure that is applied normal timing parameters so that it yields no bit-flips.
The data pointed to by \verb|A| is mapped to rows at the bottom that are applied tweaked timing parameters so that they can be accessed faster.
By mapping data of different criticality to different DRAM rows as in the figure, we can protect critical data while improving the access latency to non-critical data.

\subsection{\label{section:challenge}The Granularity Gap Problem}
\begin{figure}[t]
\begin{lstlisting}
struct node_t {
  int id;       // id of the node, critical
  struct node_t *r; // pointer to the right child, critical
  struct node_t *l; // pointer to the left child, critical
  double score; // score of this node, non-critical
};

int size = 1000 * sizeof(struct node_t);
struct node_t *nodes = malloc(size);
\end{lstlisting}
\caption{\label{figure:example_interleaved}
  Critical and non-critical data interleaved in a single C struct:
  it is not possible to protect the critical parts while storing the non-critical parts on approximate memory
  due to a large approximation granularity (e.g., 2~KB).
  }
\end{figure}

A challenge in using approximate memory is the gap between the approximation granularity and the granularity at which critical and non-critical data are interleaved.
We call this problem the {\it granularity gap problem}.
We say critical and non-critical data are {\it interleaved} when they co-locate inside one instance of a C \verb|struct| or a C++ \verb|class|.
Figure~\ref{figure:example_interleaved} shows an example of interleaved critical and non-critical data.
The data structure \verb|struct node_t| contains both critical and non-critical data, and a pointer named \verb|nodes| points to an array of \verb|struct node_t|.
To gain benefit from approximate memory for this code, we must protect the critical parts (\verb|id|, \verb|r|, and \verb|l|) while storing the non-critical part (\verb|score|) into approximate memory.
This is not possible because the approximation granularity is as larger as a few kilo bytes (say 2~KB),
while we need to enable or disable approximation with a granularity of 4 bytes to achieve it.

The granularity gap problem has been overlooked by the research community because it is not relevant to applications that have large chunks of non-critical data.
For example for deep learning applications, the non-critical data are matrices storing the weights of a network whose sizes range from a few kilo bytes to hundreds of mega bytes.
In this case, we can store entire matrices into approximate memory and the approximate granularity is not an issue.

{\bf The goal of this paper} is to prove the significance of the granularity gap problem with quantitative evidence.
First, we show that there are many applications that potentially suffer from this problem.
Second, more importantly, we show that avoiding this problem with a known technique has negative performance impact that is as large as almost canceling the benefit of approximate memory.

\section{Source Code Analysis}
To show that many real applications can potentially suffer from the granularity gap problem, we analyze source code of widely used benchmarks in this section.

\subsection{Analysis Methodology}
For a given application, we find if the data structure that can obtain benefit from approximate memory has critical and non-critical data interleaved with smaller granularity than the approximation granularity.
Because approximate memory is the most effective when an application's data that incur many cache misses are stored on it,
we focus our analysis on a data structure that incurs the largest number of cache misses within an application.
We refer to such a data structure as {\it the most cache-unfriendly data structure}.
After finding such a data structure, we analyze it to estimate if the application potentially suffers from the granularity gap problem.

To find the most cache-unfriendly data structure of an application,
we first measure the number of cache misses {\bf per instruction} using Precise Event Based Sampling (PEBS) on Intel CPUs.
PEBS is an enhancement of normal performance counters that uses designated hardware for sampling to reduce the skid between the time an event (e.g., a cache miss) occurs and the time it is recorded~\cite{Bakhvalov2018,Weaver2016}.
The small skid enables pinpointing which instruction in an application binary causes many hardware events.
We execute a benchmark with its sample dataset using linux \verb|perf|,
and the actual command line is `\verb|perf record -e r20D1:pp -- benchmark|'.
The parameter \verb|r20D1:pp| specifies a performance event whose {\it event number} is 0xD1 and the {\it umask value} is 0x20
and ``{\it counts retired load instructions with at least one uop that missed in the L3 cache}'' (described in Table 19.3 of ~\cite{intel_manual}).
Note that the ``L3 cache'' is the last level in the cache hierarchy in the CPU we use (described in Table~\ref{table:env}).
The parameter \verb|benchmark| is replaced by an actual command line to execute each benchmark.

\begin{figure}[t]
  \centering
  \includegraphics[width=0.8\columnwidth]{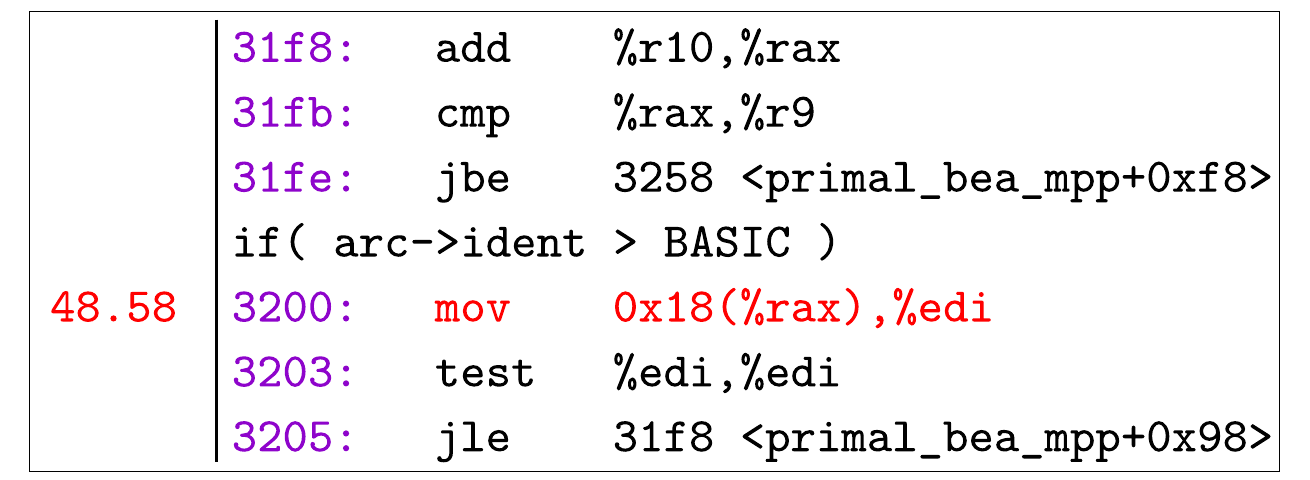}
  \caption{\label{figure:pebs_sample}Sample output of perf report: It shows an instruction, the offset of the instruction from the head of the binary, and the percentage of cache misses that it incurs (if any), from right to left. The C code, if ( arc->ident > BASIC ), corresponds to the assembly code below it.}
\end{figure}

After measuring the number of per-instruction cache misses, we find the data structure accessed by this instruction, which is the most cache-unfriendly data structure of this application.
Due to the lack of off-the-shelf tools to disassemble an arbitrary binary into C/C++ source code, we rely on human knowledge and labor to do this.
Figure~\ref{figure:pebs_sample} shows an output of \verb|perf report|, executed after a measurement by \verb|perf record|.
The measurement is done for a benchmark called \verb|mcf| in SPEC CPU 2006 and the details of the benchmarks we analyze are described in Section~\ref{section:benchmarks}.
Each line shows, from right to left, an instruction, the offset of the instruction from the head of the binary, and the percentage of cache misses it incurs (if any) against the overall cache misses in the measurement.
The C code, \verb|if( arc->ident > BASIC )|, corresponds to the lines of assembly code below it.
From the figure, we can see that the \verb|mov| instruction at offset \verb|0x3200| incurs 48.58~\% of all cache misses of this application.
We can confirm that this instruction incurs the largest number of cache misses by checking that no other instruction incurs more than this percentage.

To finding the data structure that the \verb|mov| instruction accesses given Figure~\ref{figure:pebs_sample},
we analyze the assembly code with help of the debug information and the source code.
In Figure~\ref{figure:pebs_sample}, we can see a typical pattern of assembly code where a jump instruction (\verb|jle|) follows after a compare instruction (\verb|test|, which is commonly used to compare a register with 0).
Therefore, we can guess that the \verb|mov| instruction copies a value to be compared with 0.
From the C code corresponding to this block of assembly, we can see that the value compared with 0 should be \verb|arc->ident|.
This is confirmed by the fact that \verb|BASIC| is a compile-time constant whose value is 0, and that the offset of the \verb|ident| member inside \verb|arc| is 0x18.
As a conclusion, the \verb|mov| instruction accesses the variable named \verb|arc|, whose data type is \verb|struct arc|.
Note that the same methodology is applicable to a template function in C++ as well because there is an independent piece of assembly code for each instantiation of it (i.e., no type-ambiguity exists in assembly).

\subsection{\label{section:benchmarks}Experimental Setup}
\begin{table}[t]
\centering
\caption{\label{table:benchmark}Analyzed Benchmarks}
\begin{tabular}{|c|c|c|}
\hline
\multicolumn{3}{|c|}{{\bf SPEC CPU 2006}} \\ \hline  
Name & Domain & Cache Miss Rate \\ \hline
milc & quantum simulation & 82.6\% \\
sjeng & game AI (chess) & 74.5\% \\
libquantum & quantum computing & 54.6\% \\
lbm & fluid dynamics & 49.2\% \\
omnetpp & discrete event simulation & 47.9\% \\
soplex & linear programming & 41.2\% \\
gobmk & game AI (go) & 38.4\%\\
gcc & c compiler & 36.8\% \\
mcf & optimization & 33.7\% \\
dealII & finite element analysis & 33.6\% \\
namd & molecular dynamics & 21.0\% \\ \hline \hline
\multicolumn{3}{|c|}{{\bf SPEC CPU 2017}} \\ \hline  
Name & Domain & Cache Miss Rate \\ \hline
deepsjeng\_r & game AI (chess) & 77.5 \% \\
nab\_r       & molecular modeling & 64.9 \% \\
omnetpp\_r   & discrete event simulation & 56.1 \% \\
namd\_r      & molecular dynamics & 50.4 \% \\
lbm\_r       & fluid dynamic & 48.8 \% \\
x264\_r      & video encoding & 47.3 \% \\
mcf\_r       & optimization & 43.5 \% \\
gcc\_r       & c compiler & 36.6 \% \\
blender\_r   & image processing & 35.0 \% \\
xz\_r        & data compression & 31.6 \% \\
perlbench\_r & perl interpreter & 21.4 \% \\
\hline
\end{tabular}
\end{table}

\begin{table}[t]
\centering
\caption{\label{table:env}Experiment Environment}
\begin{tabular}{|c|c|}
\hline
CPU & Intel Xeon Silver 4108 (Skylake, 8 cores) \\
Memory & DDR4-2666, 96 GB (8GB $\times$ 12)\\
LLC & 11 MB (shared across all the cores) \\
OS & Debian GNU/Linux 10 (kernel: 4.19.0-6-amd64)\\
gcc/g++ & 8.3.0 (Debian 8.3.0-6)\\
\hline
\end{tabular}
\end{table}

Table~\ref{table:benchmark} describes the benchmarks we analyze\footnote{Although deepsjeng is named `deep', it uses a classical tree search algorithm.}.
Each line shows a benchmark's name, its domain, and the cache miss rate measured by the linux \verb|perf| tool.
From both SPEC CPU 2006 and 2017, we analyze benchmarks whose cache miss rates are more than 20 \%.
We exclude others because approximate memory is not beneficial for CPU intensive benchmarks with low cache miss rates.
We also exclude ones written in Fortran because the memory layout conversion technique we discuss in Section~\ref{section:conversion} is mainly researched for programs written in C.
We include ones written in C++ because the difference between C++ and C (the existence of classes, templates, and some new syntax) do not affect the applicability of the memory layout conversion technique.

Table~\ref{table:env} shows the machine we use to execute the benchmarks.
For the input data set, we use the largest ones among provided.
That is, we use the one named \verb|ref| for SPEC CPU 2006 and the one named \verb|refrate| for SPEC CPU 2017.
The LLC miss rate is measured using the linux \verb|perf| tool with the following command: \verb|perf -e cache-misses,cache-references -- benchmark|,
where \verb|benchmark| is replaced by an actual command for each benchmark.

\subsection{Results}
\begin{table}[!t]
\centering
\caption{\label{table:results}Results of Source Code Analysis (S: is a C struct or a C++ class, P: has a pointer, F: has a fp, I: has an integer)}
\begin{tabular}{|c|c|c|c|c|c|}
\hline
\multicolumn{6}{|c|}{{\bf SPEC CPU 2006}} \\ \hline  
Benchmark    & Data Type      & S & P & F & I \\ \hline
milc         & complex[] & \Checkmark & & \Checkmark & \\
sjeng        & QTType[]  & \Checkmark & & & \Checkmark \\
libquantum   & quantum\_reg\_node\_struct[] & \Checkmark & &  \Checkmark &  \Checkmark \\
lbm          & double[]         & & & & \\
omnetpp      & cChannel & \Checkmark &  \Checkmark & &  \Checkmark \\
soplex       & Element[] & \Checkmark & & & \Checkmark \\
gobmk        & hashnode\_t[] & \Checkmark & \Checkmark & & \Checkmark\\
gcc          & rtx\_def & \Checkmark & & & \Checkmark \\
mcf          & arc[] & \Checkmark & \Checkmark & & \Checkmark \\
dealII       & double[]          & & & & \\
namd         & CompAtom[] & \Checkmark & & \Checkmark & \Checkmark \\
\hline\hline
\multicolumn{6}{|c|}{{\bf SPEC CPU 2017}} \\ \hline
Benchmark    & Data Type & S & P & F & I \\ \hline
deepsjeng\_r & ttentrty\_t[] & \Checkmark & & & \Checkmark \\
nab\_r       & INT\_T[]      & & & & \\
omnetpp\_r   & sVector    & \Checkmark & \Checkmark& \Checkmark & \Checkmark \\
namd\_r      & CompAtom[]   & \Checkmark & & \Checkmark & \Checkmark \\
lbm\_r       & double[] & & & & \\
x264\_r      & uint8\_t[]    & & & & \\
mcf\_r       & arc[]     & \Checkmark & \Checkmark & & \Checkmark \\
gcc\_r       & -          & & & & \\
blender\_r   & VlakRen[]    & \Checkmark & \Checkmark & \Checkmark & \Checkmark\\
xz\_r        & uint8\_t[], uint32\_t[]& & & & \\
perlbench\_r & char[]       &  & & & \\
\hline
\end{tabular}
\end{table}

Table~\ref{table:results} shows the analysis results.
Each row shows a benchmark, the most cache-unfriendly data structure, flags that represent the kinds of members that the data structure contains:
\begin{itemize}
\item {\bf S}: the data is either a C \verb|struct| or a C++ \verb|class|.
\item {\bf P}: the data structure contains a pointer.
\item {\bf F}: the data structure contains a floating pointer number.
\item {\bf I}: the data structure contains an integer.
\end{itemize}
The data type column is denoted by \verb|[]| if the data is managed as an array of that data type.
We regard any type compatible with an integer (e.g., \verb|char|, \verb|long|) as an integer.
If a class inherits other classes, we include the members of the parent classes as well because an instance of a child class in the memory contains all members of the parent classes.
We exclude static members and member functions because they are not stored in the memory region allocated for each instance.
We do not show the result for \verb|gcc_r| in because cache misses are scattered across many instructions.
Two data types are shown for \verb|xz_r| because two instructions incur almost the same number of cache misses.
For all the benchmarks, the instruction that incurs the largest number of cache misses existed in their own code and not in any standard C/C++ libraries.

The results show that many applications potentially suffer from the granularity gap problem.
The most cache-unfriendly data structure is either a C \verb|struct| or a C++ \verb|class| in 9 out of 11 benchmarks in SPEC CPU 2006 and 5 out of 11 benchmarks in SPEC CPU 2017.
Although there are only two benchmarks (\verb|omnetpp_r| and \verb|blender_r|) that have a pointer and a floating point number in its most cache-unfriendly data structure,
this does not mean that these two are the only benchmarks that suffer from the granularity gap problem.
For example, the data type \verb|arc| in \verb|mcf| and \verb|mcf_r| contains a pointer and an integer named \verb|cost|, which represents the cost of a graph edge.
Our previous work~\cite{Akiyama2019} shows that even if some bit-flips are injected into the member \verb|cost|, \verb|mcf| can yield the same result as an error-free execution.
Therefore, we conclude that these 14 applications ``potentially'' suffer from the granularity gap problem.

\subsection{Drawbacks of the Methodology}
Manual effort to find the data type accessed by a given instruction incurs a scalability issue and increases the chances of analysis errors.
There are two error patterns stemming from the manual effort:
\begin{enumerate}
\item Mis-identifying the variable in the source code that corresponds to a given memory access instruction
\item Mis-identifying the type of data that is stored in the identified variable in source code
\end{enumerate}
Pattern (1) can happen when the application binary has complex data/control flows for example with multiple levels of indirection (e.g., a->b->c) or
when the binary does not look similar to the source code due to compiler optimizations.
Pattern (2) can happen when the declared type of a source variable and the type of actual data stored in it are different (i.e., polymorphism).
Developing compiler support to reduce the possibilities of these errors is future work.

Another concern for our analysis arises when a member variable of a C struct or a C++ class is passed to a function by reference.
For example in Figure~\ref{figure:threat}, the same function (\verb|f|) is called either by passing \verb|&s1.v| or \verb|&s2.v| as its argument.
Finding the data type that the memory region pointed to by \verb|fp| belongs to requires an investigation of stack traces and points-to analysis~\cite{Steensgaard96}.
Although it seems more natural for a function to take a pointer of a whole \verb|struct| such as `\verb|void g(struct S1 *sp)|',
this may appear in some cases such as when a library function returns the result through a pointer.
However, we did not hit this case in any of the benchmarks in our experiment.

\begin{figure}[t]
\begin{lstlisting}
struct S1 {
  double v; // non-critical
  double vv; // non-critical
} s1;

struct S2 {
  double v; // non-critical
  int *p; // critical
} s2;

void f(double *fp) {  /* do something */ }

f(&s1.v); // (1): invoke f by passing s1.v by reference
f(&s2.v); // (2): invoke f by passing s2.v by reference
\end{lstlisting}
\caption{\label{figure:threat}Calling the same function by passing members of different structs by reference.
Identifying the data type that *fp belongs to requires stack traces and points-to analysis.
}
\end{figure}

\section{Memory Layout Conversion}
\label{section:conversion}
This section discusses an applicability of a memory layout conversion technique to avoid the granularity gap problem,
and points out that it can degrade the performance for some applications.
We show in Section~\ref{section:performance_impact} that this performance overhead is as large as almost canceling the benefit of approximate memory in some cases.

\subsection{AoS to SoA Conversion}
\begin{figure}[t]
\begin{lstlisting}
struct {
  double x;
  double y;
} points[N];

// calculate the center
double center_x = 0, center_y = 0;
for(i = 0; i<N; i++) {
  center_x += points[i].x / N;
  center_y += points[i].y / N;
}
\end{lstlisting}
\caption{\label{figure:AoS}Example of an array of structures. The data structure \{x, y\} consists an array of structures named ``points''.}
\end{figure}

\begin{figure}[t]
\begin{lstlisting}
struct {
  double x[N];
  double y[N];
} points;

// calculate the center
double center_x = 0, center_y = 0;
for(i = 0; i<N; i++) {
  center_x += points.x[i] / N;
  center_y += points.y[i] / N;
}
\end{lstlisting}
\caption{\label{figure:SoA}AoS to SoA conversion is applied to the code in Figure~\ref{figure:AoS}.
  Each member, x and y, of the struct is allocated a distinct array for it.
}
\end{figure}

An array of structures (AoS) can be converted into a structure of arrays (SoA) without changing the results of an application.
Given an array of C \verb|struct| instances, this technique converts the memory layout of an application so that each member of the C \verb|struct| is stored as a distinct array.
Figure~\ref{figure:AoS} and Figure~\ref{figure:SoA} show an example of this conversion done explicitly by hand.
The code in Figure~\ref{figure:AoS} calculates the center of $N$ points (in some sense) that are stored in memory as an array of structures.
Figure~\ref{figure:SoA} shows the converted version of the code that does the same calculation.
This version manages each member of the data structure, \verb|x| and \verb|y|, as a distinct array.
Note that following code that access the data are also changed in Figure~\ref{figure:SoA} (e.g., from \verb|points[i].x| to \verb|points.x[i]|).

Altough the AoS to SoA conversion seems very difficult at a glance for realistic applications,
existing research has proven it to be possible at compile time~\cite{Peng2007,Curial2008,Jin2010}.
The main difficulty stems from the fact that a pointer can have an arbitrary address in C/C++.
For example in Figure~\ref{figure:AoS}, if another pointer points somewhere inside a memory range pointed to by \verb|points|, it is not easy to apply the conversion without changing the application's output.
However, points-to analysis~\cite{Steensgaard96} solves this problem in almost linear time of the source code length.

Besides the technical difficulties that have been tackled by many researchers (e.g., how to apply it dynamically to programs without the source code,
how to ensure safety in weakly typed languages), a fundamental limitation of the AoS to SoA conversion is that
there is no method to precisely predict its effect on performance.
Petrank~{\it et al.}~\cite{Petrank2002} show that predicting the number of cache misses that a given data layout generates for an arbitrary memory access pattern
is NP regarding the number of data objects.
This means that one must either do exhaustive experiments for memory access patterns under interest
or use heuristics to informally estimate the performance implication.
This limitaion leads us to do the former to evaluate its performance overhead in a later section.

\subsection{Pros: Mitigate the Granularity Gap Problem}
The AoS to SoA conversion enables using approximate memory even when critical and non-critical data are interleaved by avoiding the granularity gap problem.
Because each member of the converted data structure is stored in a distinct array,
it can be mapped to a designated DRAM row that has the appropriate timing parameter for the criticality of that member.

\begin{figure}[t]
  \centering
  \includegraphics[width=\columnwidth]{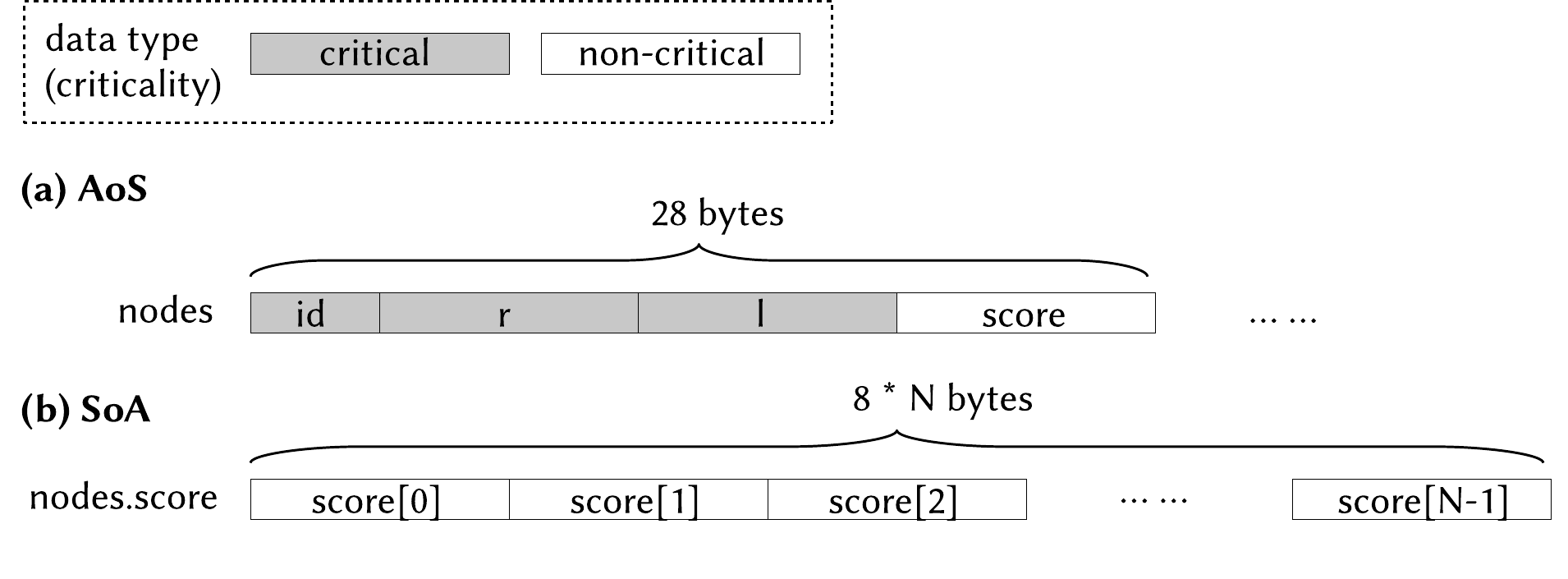}
  \caption{\label{figure:conversion_pros}The change of memory layout when the AoS to SoA conversion is applied to the code in Figure~\ref{figure:example_interleaved}.}
\end{figure}

Figure~\ref{figure:conversion_pros} depicts how we can selectively store non-critical data of the code in Figure~\ref{figure:example_interleaved} to approximate memory by the AoS to SoA conversion.
Gray boxes in the figure show critical data and white boxes show non-critical data.
In the original code that manages the data as an AoS, it is not possible to selectively protect the critical data while accelerating accesses to the non-critical data because of the granularity gap problem (Figure~\ref{figure:conversion_pros} (a)).
In the converted code that manages the data as a SoA, the non-critical data (\verb|score|) consists a distinct array and it can be mapped directly to approximate memory,
while the critical data (\verb|id|, \verb|r|, \verb|l|) can be mapped to normal memory (Figure~\ref{figure:conversion_pros} (b)).

\subsection{\label{section:conversion_cons}Cons: Negative Impact on Performance}
The disadvantage of the AoS to SoA conversion is that it can degrade the performance due to increased number of cache misses.
In the code in Figure~\ref{figure:example_interleaved}, it is highly possible that all the members of the same \verb|struct| instance (that is, for any \verb|i|, \verb|nodes[i].id|, \verb|nodes[i].r|, \verb|nodes[i].l|, and \verb|nodes[i].score|) share the same cache line.
Thus, accessing more than two members of the same \verb|struct| instance closely in time incurs at most 1 cache miss.
However, if we apply the AoS to SoA conversion to the same code, members that are in the same \verb|struct| instance in the original code do not share the same cache line.
This might increase the number of cache misses and degrade the performance depending on the memory access pattern to the data to be converted.

\begin{figure}[t]
\begin{lstlisting}
// points to the first node
struct node_t *node = malloc(sizeof(node_t) * 1000);

while(/* until some condition is met */) {
  // do something, then traverse the next node
  if (node->score > threshold)
    node = node->l;
  else
    node = node->r;
}
\end{lstlisting}
\caption{\label{figure:random_access_pattern}A sample code accessing an AoS. The definitions of node\_t is the same as Figure~\ref{figure:example_interleaved}. Applying the AoS to SoA conversion to it increases the number of cache misses.}
\end{figure}

For example, the code in Figure~\ref{figure:random_access_pattern} decides which child (either right or left) of the current node to traverse next depending on the score of the current node, and its memory access pattern is unpredictable.
When the AoS to SoA conversion is applied to this code, \verb|node->score| is stored in a different cache line from \verb|node->l| and \verb|node->r|.
Because the memory access pattern is unpredictable, an access to a new cache line incurs a cache miss every single time if the total amount of the data is large enough compared to the cache size.
Therefore, applying the AoS to SoA conversion to this code increases the number of cache misses from 1 miss per each \verb|while(...)| iteration to 2 misses per iteration.

\section{Evaluation of Performance Impact}
{\label{section:performance_impact}
The negative performance impact of the AoS to SoA conversion (the details in Section~\ref{section:conversion_cons}) is a serious concern if it cancels or even outperforms the benefit of approximate memory.
However, to the best of our knowledge, there is no study on how the AoS to SoA conversion slows down applications, because research have focused on how to speed them up.
This section introduces a new methodology to quantitatively analyze the slowdown given by the AoS to SoA conversion,
and shows that it is as large as almost canceling the benefit of approximate memory in the worst case.

\subsection{\label{section:evaluation_overview}Overview}
In order to quantitatively analyze the slowdown and show its significance, we propose a method to estimate the effect of memory layout changes incurred by the AoS to SoA conversion.
The main idea is to use a cycle accurate simulator of CPUs to run applications and reproduce the memory layout that the AoS to SoA conversion would generate inside the simulator.
The use of cycle accurate simulator has two advantages (more details are described in Section~\ref{section:simulator_discussions}):
\begin{enumerate}
\item It can quantitatively tell how much slowdown an application experiences due to memory layout conversion.
  This is important because the significance of the granularity gap problem is determined by how the slowdown is large or small relative to the benefit of approximate memory.
\item It is more robust than actually applying the conversion because it does not require complex source code analysis.
\end{enumerate}

\subsection{Pseudo Conversion by CPU Simulator}
\begin{figure*}[t]
  \centering
  \includegraphics[width=1.8\columnwidth]{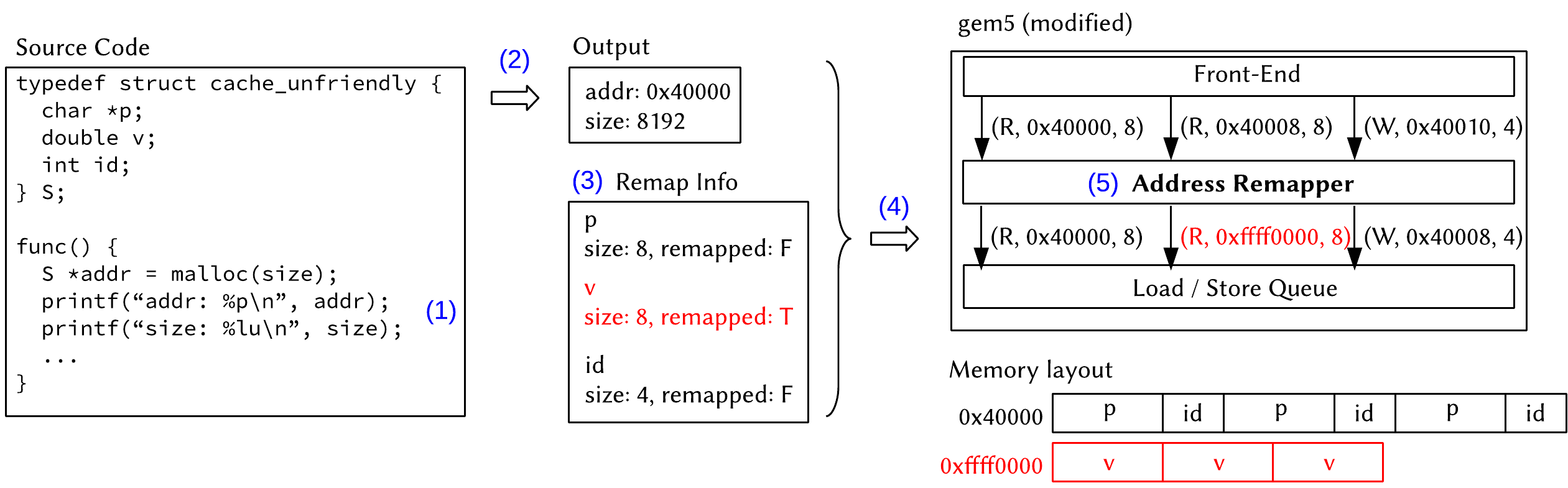}
  \caption{\label{figure:remapping}
    Simulation framework to estimate the negative performance impact of the AoS to SoA conversion.
  }
\end{figure*}

Figure~\ref{figure:remapping} shows how our simulator estimates the performance impact by reproducing the memory layout changes:
\begin{enumerate}
\item The source code of the target application is annotated so that it prints the starting addresses and the sizes of memory regions that contain the most cache-unfriendly data structure.
  For benchmarks written in C, this is done by finding \verb|malloc| calls whose return values are casted to the pointer type of that data structure.
  For benchmarks written in C++ and use the standard template libraries (STLs), this is done by replacing the memory allocator of the STLs.
\item The target application is executed on a vanilla simulator to gain the starting addresses and the sizes printed by the annotations added in step (1).
\item The {\it remap info} that decides which members of the \verb|struct| are stored in distinct arrays is defined.
  The remap info contains the size of each \verb|struct| member and a boolean value that represents if it is stored into a distinct array (we say that a member is {\it remapped} if this value is \verb|true|).
\item A simulation is started on our modified simulator with information obtained in step (2) and step (3) (the starting addresses and sizes of memory regions and the remap info) passed as inputs.
\item While in a simulation, the target addresses of memory access instructions are investigated.
  If the target address points to a remapped member, it is converted to reproduce the memory layout that the AoS to SoA conversion would generate.
\end{enumerate}

The address conversion is done when the front-end of the CPU inserts requests into the load store queue.
The component that converts addresses is illustrated as the {\it address remapper} in Figure~\ref{figure:remapping}.
This is because the border between the front-end and the load store queue is a place right after an accessed address is determined and right before it is used.
Inside the front-end, the target address of a memory access instruction might not be ready, for example when the register that contains the address is an operand of a not-yet-committed instruction.
Inside the load store queue, the address of a request is used to access the caches first before accessing the main memory.
Therefore, we convert an address before it is referenced in the load store queue to maintain the cache consistency.

Three requests are passed from the front-end to the address remapper in Figure~\ref{figure:remapping}:
\begin{enumerate}
\item 8-byte read request to \verb|0x40000|.
\item 8-byte write request to \verb|0x40008|.
\item 4-byte read request to \verb|0x40010|.
\end{enumerate}
From the starting address of the memory region that contains the most cache-unfriendly data structure and the remap info,
the address remapper can find that the first request reads the member \verb|p|.
Because the remap info specifies that \verb|p| is not remapped, the request is passed as-is to the load store queue.
The second request accesses the member \verb|v|.
Because the remap info specifies that it is remapped, its target address is converted into an unused address (\verb|0xffff0000| in the figure).
The third request accesses the member \verb|id|.
Although it is not remapped, its address is shifted by 8 bytes because the previous member \verb|v| is remapped ``away''.
Thus, the target address is converted to \verb|0x40008|.
As a result, the memory layout from the application's point of view is converted into the one shown in the figure.
The member \verb|v| consists a distinct array and the other members are packed as if there is no \verb|v| in-between.

\subsection{\label{section:simulator_discussions}Discussions}
With regard to the effect of memory layout conversion, there are efforts to estimate how it speeds up applications~\cite{Peng2007,Miucin2018,Ye2019} by investigating their memory access traces without applying the memory layout changes themselves.
They measure the access frequencies to \verb|struct| members and the access affinities between them from a memory trace of unmodified source code.
Given these metrics, they suggest which members should be placed closer in memory and which members should be separated to a different memory region.
However, we cannot directly leverage this method for our purpose because they only ``suggest'' better memory layouts but do not quantitatively estimate the performance impact of the suggested layouts.
A difficulty when it is used for our purpose is that memory layout conversion has two effects of opposite directions:
(1) {\bf slowdown} caused by increased number of cache misses due to separation of members with strong affinities, and
(2) {\bf speedup} caused by decreased size of members that are not separated as distinct arrays.
For example, the size of \verb|arc| data structure in \verb|mcf| is 72 bytes (9 members $\times$ 8 bytes/member) and separating one member as a distinct array makes the size of the rest to 64 bytes, which fits within a cache line.
In fact, we tried our best but failed to predict the results in Section~\ref{section:result_conversion} using the same access affinity metric as existing work~\cite{Miucin2018,Ye2019} (we omit the details for brevity).

A challenge of actually applying the SoA to AoS conversion in the compiler-level stems from the fact that pointers can contain any addresses in C/C++ and the values held by pointers cannot be decided by a static analysis.
Due to this, although theoretically possible by points-to analysis~\cite{Steensgaard96}, it is not easy to robustly implement the AoS to SoA conversion in the compiler level.
Some old \verb|gcc| versions supported structure reordering, which reorders members of a C \verb|struct| and requires the same type of analysis.
However, this feature was removed because it ``{\it did not always work correctly}''~\cite{gcc_4.8_release_note}.
In contrast, because our method converts memory addresses inside a simulator at runtime, there is no difficulty in finding the address that a pointer contains.

A disadvantage of our method is that a cycle accurate simulation is needed for every single conversion pattern.
This is not always possible because the number of memory layout conversion patterns increases exponentially to the number of members in the most cache-unfriendly data structure.
On the other hand, if can we somehow estimate the slowdown only from access frequencies and affinities, we can estimate the slowdown of all conversion patterns at once
because the access frequencies and affinities can be obtained by one execution of a non-modified application.

\subsection{Experimental Setup}
\begin{table}[t]
\begin{center}
\caption{\label{table:simulated_machine}Simulated Environment}
\begin{tabular}{c|c}
  \hline
  ISA & x86\_64 \\
  Frequency & 3 GHz \\
  Issue Width & 8 \\
  Reorder Buffer & 192 entries \\
  L1 cache & 32 KB, 2 way, 32 MSHRs, 2 cycles/miss\\
  L2 cache & 2 MB, 8 way, 32 MSHRs 20 cycles/miss\\
  Mem Ctrl Latency & 75 ns \\
  \hline
\end{tabular}
\end{center}
\end{table}

Table~\ref{table:simulated_machine} shows the simulated environment.
The ``Mem Ctrl Latency'' shows the length of time between a point when the CPU sends a request to the memory controller and a point when it receives the response.
The memory access latency from software point of view additionally contains the time it takes to miss the caches, which is  7.3 ns (= (2 + 20) cycles $\times$ $\frac{1}{3}$ ns per cycle) and makes up a total of 82.3 ns.
We use version 20.0.0.0 (the latest version as of May 2020) of gem5 and its SE mode.
Besides simulating instructions, this mode {\bf emulates} system calls by replacing them with calls to normal functions defined in the gem5 source code.
It allows easy simulation because there is no need to run an entire OS on the simulator,
and there should be no noticeable impact on the results as non OS-intensive workloads have few system call invocations.
The benchmark binaries are compiled by gcc 8.3.0 (Debian 8.3.0-6).

\begin{figure*}[t]
  \centering
  \includegraphics[width=1.7\columnwidth]{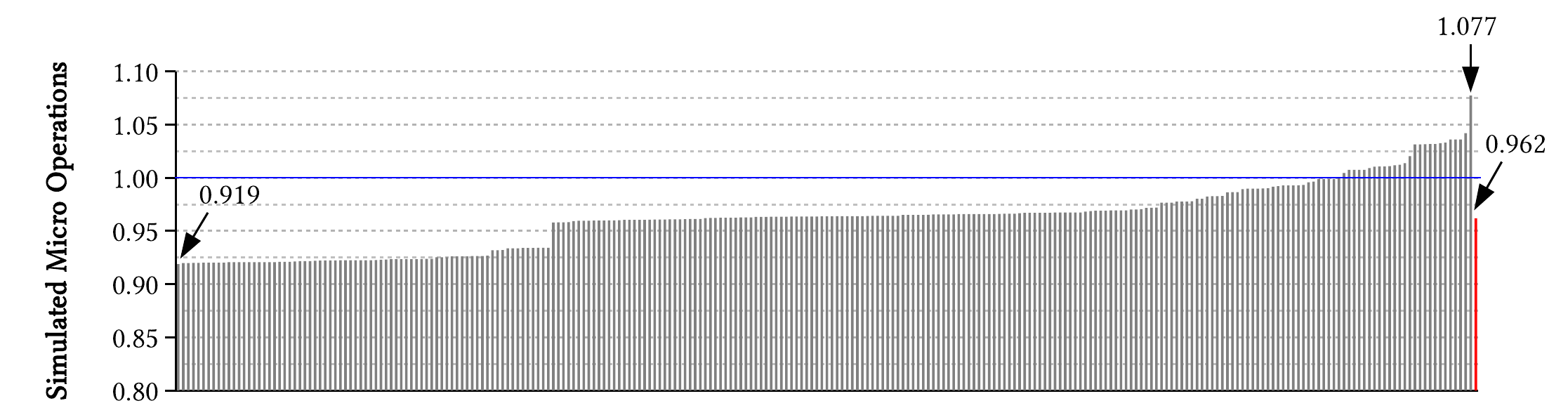}
  \caption{\label{figure:result_mcf}Evaluation result for mcf\_r
  }
\end{figure*}

\begin{figure}[t]
  \centering
  \includegraphics[width=\columnwidth]{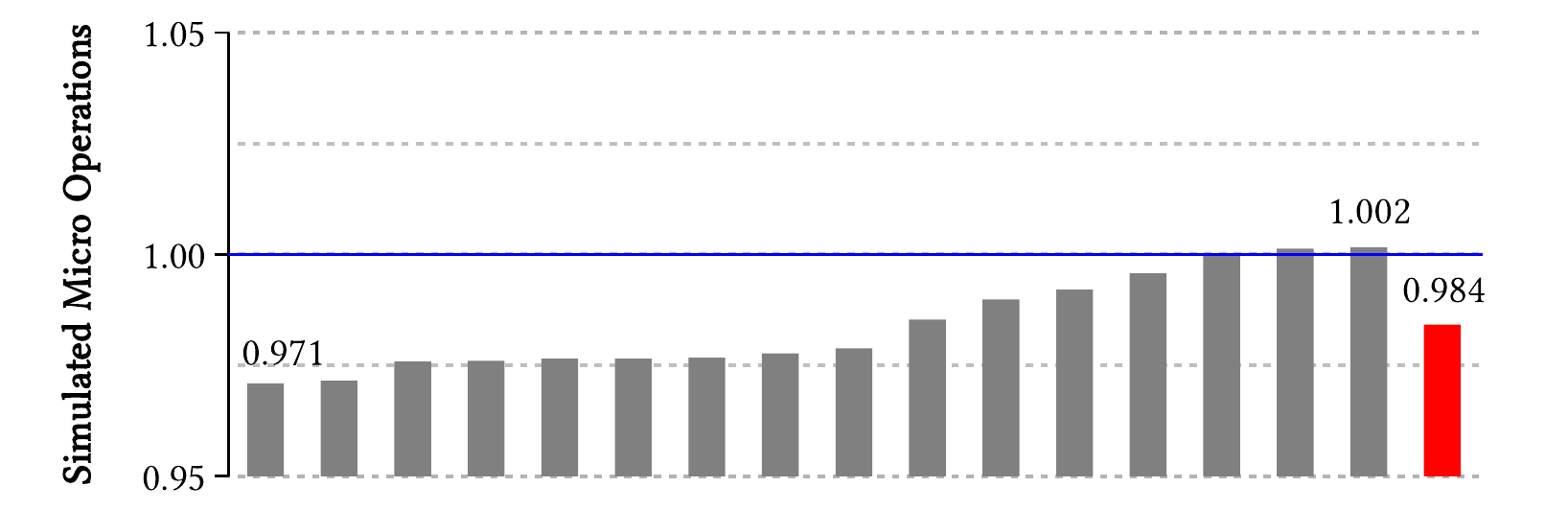}
  \caption{\label{figure:result_deepsjeng}Evaluation result for deepsjeng\_r
  }
\end{figure}

\begin{figure}[t]
  \centering
  \includegraphics[width=\columnwidth]{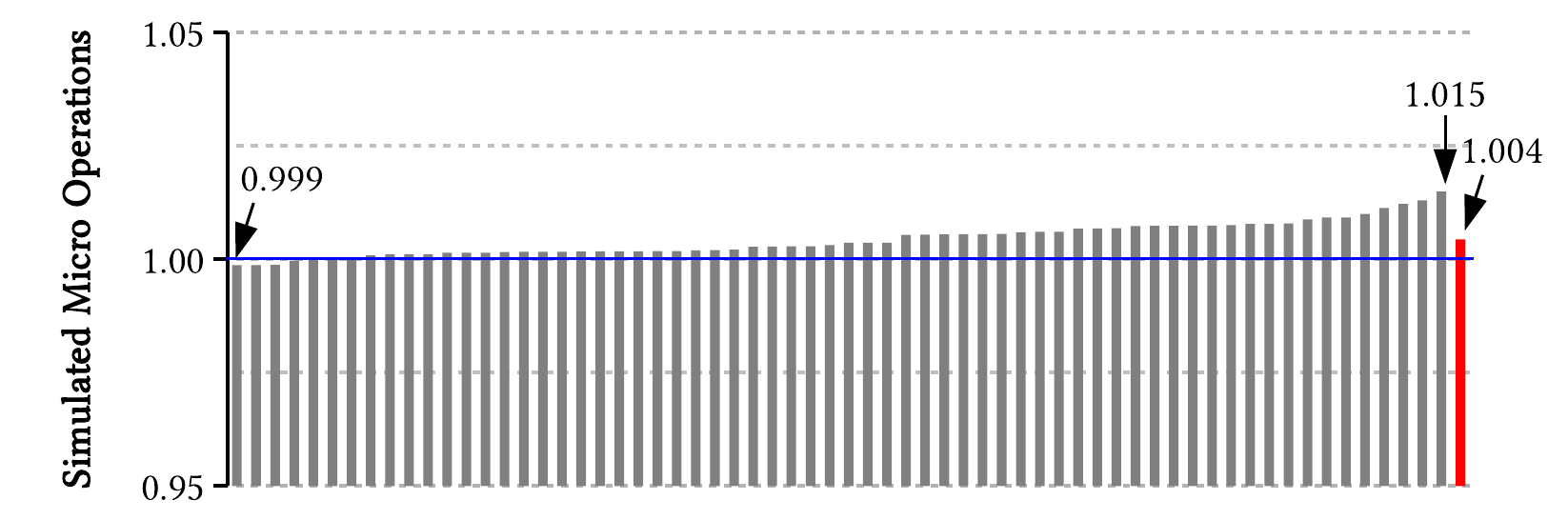}
  \caption{\label{figure:result_namd}Evaluation result for namd\_r
  }
\end{figure}

For the evaluation, we first skip the initial phase of each benchmark using a simulation mode that only {\bf emulates} a CPU using the \verb|AtomicSimpleCPU| model.
After the initialization phase, we simulate a fixed number of {\it ticks} using a mode that {\bf simulates} an out-of-order CPU using the \verb|DerivO3CPU| model.
A tick is the notion of time in gem5 and it is 1/1000 ns in our configuration.
We simulate 200 billion ticks after the initialization phase, which is equal to 0.2 seconds in the simulated world.
The initialization phase of each benchmark is determined by investigating the source code.

We evaluate three benchmarks from SPEC CPU 2017, namely \verb|mcf_r|, \verb|deepsjeng_r|, and \verb|namd_r|.
For each benchmark, we test every possible memory layout conversion pattern and compare the performance.
Let $N$ be the number of members in the most cache-unfriendly data structure in each benchmark, we test all $2^{N-1}$ cases of remapping.
Note that remapping a given $M$ members (and not remapping the rest) is equivalent to not remapping the $M$ members (and remapping the rest) with regard to the memory layout.
We exclude \verb|blender_r| and \verb|omnetpp_r| although their most cache-unfriendly data structures are C++ \verb|class|.
This is because (1) the source code of \verb|blender_r| does not have clear separation between the initialization phase and the main computation,
and (2) \verb|omnetpp_r| has 21 members in its most cache-unfriendly data structure (\verb|sVector|) and it is not possible to test $2^{20}$ possibilities.
The latter stems from a disadvantage of our method that we must conduct time-consuming simulations for different memory layout conversion patterns,
as we describe in Section~\ref{section:evaluation_overview}.

\subsection{\label{section:result_conversion}Results}
Figure~\ref{figure:result_mcf}, Figure~\ref{figure:result_deepsjeng}, and Figure~\ref{figure:result_namd} show the evaluation results of performance degradation for \verb|mcf_r|, \verb|deepsjeng_r|, and \verb|namd_r|.
Each bar corresponds to a memory layout conversion pattern and each graph has $2^{N-1} + 1$ bars, where $N$ is the number of members of the most cache-unfriendly data structure.
The right-most bar shows the average of all patterns.
The $y$ values show the number of executed micro operations during simulation normalized to the value when memory layout conversion is not applied.
The bars are ordered by their $y$ values.
Because we simulate a fixed number of ticks, lower bars show larger negative performance impact.

\begin{description}
\item[mcf\_r:] The most cache-unfriendly data structure (arc) has 9 members and there are $2^{8} = 256$ memory layout conversion patterns. 
  Among them, 229 patterns yield worse performance than the no conversion case (the $y$ values are smaller than 1).
  The lowest performance is observed when the first three members are remapped to consist distinct arrays, and its performance is 8.13 \% slower than the no conversion case.
  The 9 members are all 8 bytes in size and thus this pattern makes the size of the non-remapped part to be 48 bytes.
  The average negative performance impact is 3.81 \%.
\item[deepsjeng\_r] The most cache-unfriendly data structure (ttentry\_t) ``wraps'' an array of length 4, and each member of the array is another C \verb|struct| whose number of members is 5 (see Figure~\ref{figure:deepsjeng_data} for illustration).
  We apply the same remapping policy for the same members of the inner data structure, resulting in $2^4 = 16$ memory layout conversion patterns (if ttentry\_t.array[0].d1 is remapped, ttentry\_t.array[i].d1 are remapped for any i $\in \{1,2,3\}$).
  The lowest performance is observed when the first and the fourth member (d1 and d4 in Figure~\ref{figure:deepsjeng_data}) are remapped, resulting in 2.90~\% slowdown.
\item[namd\_r] The most cache-unfriendly data structure (CompAtom) has 7 members and there are $2^{6} = 64$ memory layout conversion patterns.
  The negative performance impact to this application is negligible (0.1~\% in the worst case).
  In average, the performance is improved by 0.4~\%.
\end{description}

\begin{figure}[t]
\begin{lstlisting}
struct inner_struct {
  type1 d1;
  type2 d2;
  type3 d3;
  type4 d4;
  type5 d5;
};

struct ttentry_t {
  struct inner_struct array[4];
};
\end{lstlisting}
\caption{\label{figure:deepsjeng_data}
  The most cache-unfriendly data structure (ttentry\_t) of deepsjeng\_r.
  The details of inner\_struct are not shown because SPEC CPU is non-free software.
  }
\end{figure}

The negative performance impact of the memory layout conversion to avoid the granularity gap problem is not negligible compared to the benefit of approximate memory.
For example, Kim~{\it et al.}~\cite{Kim2018} report that the average speedup of SPEC CPU 2006 benchmarks when the timing constraints are violated is around 4 - 5~\% (Figure 8 of~\cite{Kim2018}).
Note that their system, Solar-DRAM, does not reduce the latency to the extent that bit-flips are visible to the applications.
Even if we assume that the performance gain by approximate memory (allowing bit-flips to be visible to the application) is twice as large as Solar-DRAM, it is almost canceled in the worst case by the performance overhead due to the granularity gap problem (8 - 10~\% speedup vs. 8.13~\% slowdown).
Another research by Tovletoglou~{\it et al.}~\cite{Tovletoglou2020} report that their system can save up to around 12.5~\% of overall (CPU + memory) energy consumption  for \verb|mcf| in SPEC CPU 2006 (the bar labeled as ``429'' in Figure 7~(c) of~\cite{Tovletoglou2020}) by approximate memory (prolonging \verb|tREF|).
Prolonging \verb|tREF| reduces not only the memory access latency but also the energy consumption of DRAM~\cite{Das2018}, which is another benefit of approximate memory besides performance.
If we assume that the negative performance of memory layout conversion to \verb|mcf| is similar to the one to \verb|mcf_r|\footnote{They are quite similar and the most cache-unfriendly data stuructures are the same.}, the 12.5~\% gain is deducted by a non-negligible amount, because we have up to 8.13~\% performance overhead by memory layout conversion.
Therefore, we conclude that the granularity gap problem is a significant issue and the research community needs efforts to solve it with low overhead
to expand the benefit of approximate memory into wider range of applications.

\section{Related Work}
To the best of our knowledge, we are the first to study the granularity gap problem.
One of the reasons is that it is not relevant when we consider storing only large arrays of numbers such as weight matrices of a neural network to approximate memory.
However, as we point out in this paper, it is a significant problem for many realistic applications.
Esmaeilzadeh~{\it et al.} mention~\cite{Esmaeilzadeh2012} about this problem a bit, but they provide no further investigation.

Nguyen~{\it et al.}~\cite{Nguyen2020} propose a method that partially mitigates the granularity gap problem.
It transposes rows and columns of data layout inside DRAM so that a chunk of data is stored across many rows that have different error rates.
This enables protection of important bits (e.g., the sign bit of a floating point number) while aggressively approximating less important bits.
This mechanism is effective for DNNs because they require the whole part of a large weight matrix at once
and the number of memory accesses do not increase regardless of the data layout.
However, it is not effective in general cases where memory is accessed with smaller granularity.

Mapping data into memory regions with different error rates depending on its criticality is commonly proposed.
Liu~{\it et al.}~\cite{Liu2011} partition a DRAM bank into bins with proper refresh interval and ones with prolonged refresh interval.
Each data is store into either type of bins depending on the criticality specified by the programmer.
Although they do not discuss the minimum bin size, it cannot be smaller than a DRAM row as we discuss in this paper.
Chen~{\it et al.}~\cite{Chen2016} propose a memory controller that maps data into different DRAM banks with different error rates depending on the criticality of the data.
Because this method is bank-based, the approximation granularity is limited to the bank size.
A typical DDR3/DDR4 DIMM module has 2~GB to 16~GB with either 8 or 16 banks, resulting in a typical bank size of 256~MB to 2~GB.
Raha~{\it et al.}~\cite{Raha2017} advance a previous work~\cite{Liu2011} by measuring each bin's error rate at a given prolonged refresh interval
and assigning them to approximate data in the ascending order of the error rate.
They realize the bin size (or ``page size'' in their terminology) of 1~KB by measuring the average error rate per 1~KB.
Although this approach could be further pursued to realize smaller page sizes,
it still cannot control error rates per byte as it just measures them and use appropriate pages.

Our previous work~\cite{Akiyama2020} investigates the source code of SPEC CPU 2006 benchmarks and shows that there are many applications that potentially suffer from the granularity gap problem.
Besides adding more data for the source code analysis to further strengthen the analysis,
the novelty of this paper is that we quantitatively analyze slowdown caused by the granularity gap problem and show experimental results on some benchmarks to further understand the significance of the granularity gap problem.

\section{Conclusion}
In this paper, we investigated the granularity gap problem of approximate memory.
The problem arises due to the difference between approximation granularity and the granularity of data criticality of realistic applications.
Because the former is as large as a few kilo bytes in realistic DRAM modules and the latter is often a few bytes, we cannot map data of these applications directly on approximate memory.
We analyzed source code of SPEC CPU 2006 and 2017 benchmarks and found that 14 out of 22 benchmarks potentially suffer from this problem.
In addition, we pointed out the applicability of a memory layout conversion technique to this problem and negative performance impact of it.
We proposed a simulation framework to quantitatively analyze the negative performance impact of this technique, and found that the the performance can be degraded by up to 8.13 \% in our tested cases.
We conclude that the granularity gap problem is a significant issue and it requires more attention from the research community.

\begin{acks}
  This work was supported by JST, ACT-I Grant Number JPMJPR18U1, Japan.
\end{acks}

\bibliographystyle{ACM-Reference-Format}
\bibliography{ICPE2021}

\end{document}